\documentclass[fleqn,usenatbib]{mnras}
\usepackage{mathtools}
\usepackage{newtxtext}
\usepackage{verbatim}
\usepackage{ae,aecompl}
\usepackage{tablefootnote}
\usepackage{graphicx}	
\usepackage{amsmath}	
\usepackage{amssymb}	

\usepackage{float}
\usepackage{multirow}
\usepackage{units}
\usepackage{placeins}
\usepackage{xcolor}

\title[HFQPOs in GRS 1915+105]{Evidence of oscillating `compact' Comptonized corona in GRS 1915+105: Insights into HFQPOs with {\it AstroSat}}

\author[Harikesh et al.]{
Sreehari Harikesh,$^{1}$\thanks{E-mail: hjsreehari@gmail.com}
Seshadri Majumder,$^{2}$
Santabrata Das$^{2}$\thanks{E-mail: sbdas@iitg.ac.in}
and Anuj Nandi$^{3}$\thanks{E-mail: anuj@ursc.gov.in}
\\
$^{1}$Department of Physics, University of Haifa, Haifa 3498838, Israel\\
$^{2}$Indian Institute of Technology Guwahati,
Guwahati, 781039, India\\
$^{3}$Space Astronomy Group, ISITE Campus, U. R. Rao Satellite Centre,
Outer Ring Road, Marathahalli, Bangalore, 560037, India.
}

\date{Accepted XXX. Received YYY; in original form ZZZ}

\pubyear{2025}

\begin{document}
\label{firstpage}
\pagerange{\pageref{firstpage}--\pageref{lastpage}}
\maketitle

\begin{abstract}

    We present, for the first time, an in-depth dynamical analysis of the spectro-temporal properties of the soft variability classes ($\delta$, $\kappa$, $\omega$, and $\gamma$) of GRS 1915+105 during the detection of $\sim$70 Hz High-Frequency Quasi-periodic Oscillations (HFQPOs) using {\it AstroSat} data. The wide-band spectra ($0.7-50$ keV) are well described by thermal Comptonization along with an extended power-law component. Additionally, power spectra ($0.01-500$ Hz) indicate that Comptonized photons ($6-25$ keV) primarily contribute to the HFQPOs. Our findings reveal that high (low) count rates referred to as `non-dips' (`dips') in the light curves of the variability classes correspond to the detection (non-detection) of HFQPOs. Accumulated `non-dips' (`dips') spectra are modelled separately using thermal Comptonization (\texttt{nthComp}) as well as \texttt{kerrd} which indicates harder spectra and smaller inner disc radius during the detection of HFQPOs. We conduct dynamical analyses (every 32 s) to trace the presence of HFQPOs, and variations in thermal Comptonization parameters ($\Gamma_{\rm nth}$ and ${\rm N}_{\rm nth}$). Moreover, we observe a positive correlation of `non-dips' with QPO strength, ${\rm HR}1$, and ${\rm N}_{\rm nth}$, while $\Gamma_{\rm nth}$ shows an anti-correlation, suggesting that high-energy photons from the Comptonized corona are responsible for the HFQPOs. Furthermore, we estimate the size of the Comptonized corona using \texttt{kerrd} and \texttt{diskpn} to be $\sim 2.8 - 16$ $r_{\rm g}$. Thus, we infer that a `compact' oscillating corona likely modulates the high-energy radiation, exhibiting the $70$ Hz HFQPOs in GRS 1915$+$105.

\end{abstract}


\section{Introduction}

\label{sec:intro}

High-frequency Quasi-periodic Oscillation (HFQPO) feature is regarded as a useful tool for understanding the accretion dynamics around black hole X-ray binaries (BH-XRBs), as they carry the imprint of strong gravity from the central source. Several BH-XRBs, such as GRS 1915+105 \cite[]{Morgan1997,Strohmayer2001}, GRO J1655-40 \cite[]{Remillard1999}, XTE J1550-564 \cite[]{Homan2001}, H1743-322 \cite[]{Homan2005} and IGR J17091-3624 \cite[]{Altamirano2012} were often seen to exhibit HFQPOs in the {\it Rossi X-ray Timing Explorer} ({\it RXTE}) era. Among them, GRS 1915+105 generally exhibits various structured variability classes \cite[and references therein]{Belloni-etal00, Klein-Wolt-etal2002, Naik-etal2002, Hannikainen-etal2005}, and also demonstrates ultra-fast modulations in some of these classes \cite[]{Belloni2006, Belloni-Altamirano13}. Additionally, temporal analyses of these structured variabilities often reveal HFQPOs at $\sim70$ Hz, particularly in the softer variability classes ($e.g.$, $\delta$, $\gamma$, $\kappa$, $\mu$, and $\omega$), which is also confirmed by {\it AstroSat} \cite[]{Belloni2019, Sreehari-etal2020, Majumder2022,Majumder-etal2023}.

The wide-band energy spectra of GRS 1915+105 were satisfactorily described by the thermal disc photons and non-thermal Comptonized emissions \cite[]{Zdziarski-etal2005}. In fact, a disc-corona configuration was also suggested by \cite{Taam1997,Vilhu2001}, where they modelled the {\it RXTE} energy spectra using the multi-temperature blackbody and a \texttt{powerlaw}. \cite{Vilhu1998} studied the changes of the disc temperature, optical depth and inner disc radius of GRS 1915+105 during the rapid variations in photon count rates. Subsequently, \cite{Titarchuk2009} attempted to explain the energy spectra using the bulk motion Comptonization model that comprises both hard and soft thermal components. Meanwhile, the time-resolved evolution of the energy spectral parameters of this source was reported by \cite{Migliari2003} during its $\beta$ class variability. 

Indeed, the  origin of HFQPOs remains of great interest as these oscillations are transient as well as subtle. However, the conclusive consensus on the origin of HFQPOs is not settled yet. In the recent past, \cite{Sreehari-etal2020} speculated that the HFQPOs are manifested possibly due to the coherent modulation of the `compact' Comptonized corona at the inner part of the disc. Similar findings are corroborated in some of the variability classes observed with {\it AstroSat} \cite[and references therein]{Majumder2022}. Earlier, while studying $\gamma$ class of GRS 1915+105, \cite{Belloni2001} found HFQPOs during the low count rate (`dips') in $13 - 30$ keV energy band, although such signature disappears at low energies ($2 - 13$ keV). Meanwhile, \cite{Majumder2022} showed that when the thermal Comptonized component is dominant, HFQPOs are seen in the `softer' variability classes ($\delta,\kappa,\omega,\gamma$) of GRS 1915+105. Following this, \cite{Majumder-etal2023} observed that the soft photons lag behind the hard photons during detection of these HFQPOs. \cite{Majumder2025} associated HFQPOs with a larger covering fraction and increased Comptonization flux along with lower optical depth. However, there are lack of attempts to investigate the correlation between HFQPOs and the energy spectral parameters of GRS 1915$+$105 within structured variability classes such as $\kappa$ and $\omega$, which exhibit variations on shorter timescales of a few tens of seconds. 

Being motivated with this, in this manuscript, we aim to investigate the dynamical variation of the HFQPOs in different structured variability classes. To achieve our goal, we carry out a comprehensive analysis to examine the appearance and disappearance of HFQPOs within each variability class. While doing so, we consider {\it AstroSat} observations and make use of the short duration ($32$ s) data with reasonable photon statistics. With {\it AstroSat} data, we examine the possible correlations between the structured variabilities of the light curves and the variation of the spectral parameters to infer the origin of HFQPOs in GRS 1915$+$105.

The paper is organized as follows. In \S2, we describe the selection of observations. The modelling and results are presented in \S 3. Finally, we discuss our findings and present conclusion in \S 4.

\section{Observations}
\label{s:obs}
\begin{figure*}
		\includegraphics[width=\textwidth]{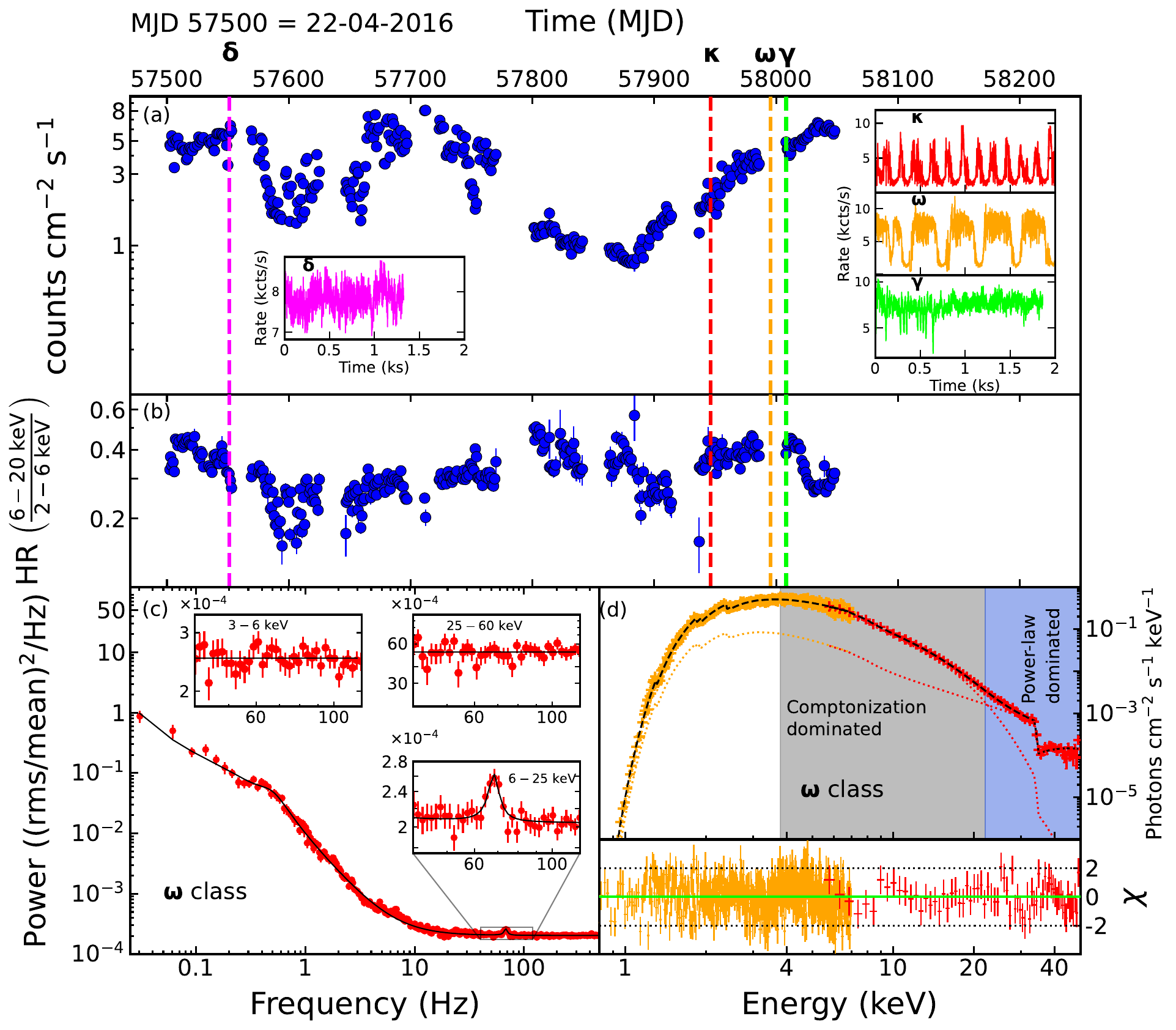}
	\caption{Variation of (a) photon count rate as observed with {\it MAXI} and (b) the hardness ratio (HR) with time. Dashed vertical lines denote different variability classes as observed by {\it AstroSat} and the corresponding light curves are shown at the inset of panel (a). (c) Power spectral distribution in rms-frequency space is plotted for $\omega$ class observation (Orbit 10394, MJD 57995.40). At the insets, PDS in different energy band are shown. The broadband spectrum of $\omega$ class modelled using \texttt{constant$\ast$Tbabs$\ast$edge$\ast$smedge(nthComp+powerlaw)} is presented in panel (d). The y-axis of all the four main panels are in logarithmic scale. See the text for details.} 
	\label{fig:PDS10394}
\end{figure*}

The extensive {\it RXTE} observations facilitated the insightful findings related to the HFQPOs of GRS 1915+105 \cite[]{Morgan1997,Cui1999,Belloni2001,Belloni-Altamirano13}. Due to the larger collecting area, {\it AstroSat/LAXPC} \cite[]{Antia-etal17} provides high quality data that allows one to carry out in-depth analyses as well as understanding of the HFQPO features. Recently, using {\it AstroSat/LAXPC} observations, HFQPOs are reported in several variability classes ($\delta$, $\kappa$, $\gamma$, and $\omega$) of GRS 1915+105 \cite[]{Belloni2019,Sreehari-etal2020,Majumder2022}. In this work, we re-analyse these `softer' variability classes between MJD 57500 and MJD 58050 (see Figure \ref{fig:PDS10394}), during which HFQPOs of $\sim 70$ Hz frequency are observed. In particular, we choose four {\it AstroSat} observations corresponding to orbit numbers 3819 ($\delta$ class), 9670 ($\kappa$ class), 10394 ($\omega$ class), and 10583 ($\gamma$ class), respectively. Among these observations, $\kappa$ and $\omega$ classes show structured variabilities with clear presence of high count rate (`non-dips') and low count rate (`dips') regions. Indeed, the light curves of this kind are ideally suited for studying the spectro-temporal properties of HFQPOs.

In order to examine light curves, power density spectra and energy spectra, we reduce {\it AstroSat} data following \cite{Agrawal-etal2018,Sreehari-etal2019,Sreehari-etal2020,Majumder2022}. For timing analysis, we combine {\it LAXPC10} and {\it LAXPC20} data, whereas only {\it LAXPC20} data are used for spectral analysis. {\it AstroSat/LAXPC} light curves are generated with $1$ s resolution for calculating hardness ratios using \texttt{LaxpcSoft}\footnote{\url{https://www.tifr.res.in/~astrosat_laxpc/LaxpcSoft.html}}. While generating the power density spectrum, we use {\it LAXPC} light curves of resolution $0.001$ s. The broadband energy spectra are generated by combining {\it SXT} ($0.7 - 7$ keV) and {\it LAXPC20} ($3 - 50$ keV) data. We also use MAXI data to study the long-term variability of the source over the duration of our interest (MJD 57500 to MJD 58100)\footnote{\url{ http://maxi.riken.jp/top/index.html}}.

\begin{table*}
        \caption{The parameters obtained from broadband ($0.7 - 50$ keV) spectral modelling (average) of the {\it AstroSat} observations of $\delta$, $\kappa$, $\omega$ and $\gamma$ classes of GRS 1915+105 are tabulated. Three different model combinations are used as Model-1: \texttt{constant$\ast$Tbabs$\ast$edge$\ast$smedge(nthComp+powerlaw)}, Model-2: \texttt{Tbabs$\ast$edge$\ast$smedge(diskpn+powerlaw)} and Model-3: \texttt{Tbabs$\ast$edge$\ast$smedge(kerrd+powerlaw)}. The bolometric X-ray luminosities are computed in $1 - 100$ keV energy band. See the text for details.}
	\resizebox{1.0\textwidth}{!}{
		\begin{tabular}{|l|c|c|c|c|c|c|c|c|c|c|c|c|c|c}
			\hline
			&&\multicolumn{4}{c|}{Model-1}&\multicolumn{3}{c|}{Model-2}&\multicolumn{3}{c|}{Model-3}&&& \\
			\hline
			MJD (Orbit)  & Class &${\rm \Gamma _{nth}}$&${\rm \Gamma}_{\rm PL}$&$\chi^2_{\rm red}$& R$_{\rm in}$ & ${\rm T}_{\rm max}$ & ${\rm \Gamma}$ &$\chi^2_{\rm red}$ & R$_{\rm in}$& ${\rm \Gamma}$ & $\chi^2_{\rm red}$&L$_{\rm bol}$ & L$^{^{\rm D}}_{\rm bol}$  & L$^{^{\rm N}}_{\rm bol}$ \\
			& ($\nu_{\rm QPO}$ Hz) && &  ($\chi^2/dof$) &(r$_{\rm g}$) &(keV)&  & ($\chi^2/dof$) &(r$_{\rm g}$)&& ($\chi^2/dof$) &(L$_{\rm Edd}$)&(L$_{\rm Edd}$)&(L$_{\rm Edd}$)\\
			\hline
			57551.04 (3819) &$\delta$ ($69.18$) &1.82 $\pm$ 0.21 & 3.15 $\pm$ 0.23 & 56/61 & $14.9 \pm 2$ & 2.77 $\pm$ 0.01 & 3.23 $\pm$ 0.01 & 54/62 & $3.72_{-0.18}^{+0.18}$ & $2.82_{-0.04}^{+0.09}$ & 0.90 (50/55) &  0.18 & --- & --- \\
			57946.34 (9670) &$\kappa$ ($70.45$)  &1.87 $\pm$ 0.05&2.14 $\pm$ 0.55&733/601 &$14.62 \pm 0.29$ & 2.59 $\pm$ 0.01 & 2.91 $\pm$ 0.01 & 716/598 &  $3.56_{-0.22}^{+0.12}$  & 3.02 $\pm$ 0.01& 1.22 (738/601) &0.10&0.07&0.15\\
			57995.40 (10394) & $\omega$ ($68.09$) &1.68 $\pm$ 0.08&2.41 $\pm$ 0.38&724/625 &$16.56 \pm 0.23$ & 2.75 $\pm$ 0.01 & 2.77 $\pm$ 0.01 & 712/628 & $3.27_{-0.09}^{+0.09}$ &2.76 $\pm$ 0.01 & 1.08 (682/630) & 0.19&0.09&0.26\\
			58008.18 (10583) & $\gamma$ ($72.32$)  &1.92 $\pm$ 0.11&2.61 $\pm$ 0.58&604/534 & $15.49 \pm 0.35$& $2.84 \pm 0.01$ & $2.93 \pm 0.01$ & 556/533 & $3.58_{-0.05}^{+0.08}$& $2.92_{-0.01}^{+0.02}$ & 1.02 (546/535) & 0.25 & --- & --- \\
			\hline
			
		\end{tabular}
	}
	
	\label{tab:spectra}
\end{table*}

\begin{table*}
	\caption{Broadband ($0.7 - 50$ keV) spectral parameters for accumulated `dips' and `non-dips' segments. Here, adopted models are as Model-1: \texttt{Tbabs$\ast$edge$\ast$smedge(nthComp+powerlaw)} and Model-3: \texttt{Tbabs$\ast$edge$\ast$smedge(kerrd+powerlaw)}. See the text for details.}
	    \resizebox{1.0\textwidth}{!}{
	    \begin{tabular}{|c|c|c|c|c|c|c|c|c|c|c|c|c|c|c|}
	    \hline
            &&\multicolumn{7}{c|}{Model-1}&\multicolumn{5}{c|}{Model-3} \\
            \hline
	    Orbit& Class & $\Gamma_{\rm nth}$&N$_{\rm nth}$ & $\Gamma_{\rm PL}$&N$_{\rm PL}$ & smedge (keV) & kT$_{\rm e}$ (keV)&($\chi^2/dof$)&R$_{\rm in}$(r$_{\rm g}$)  &N$_{\rm kerr}$& $\Gamma$ &N$_{\rm PL}$ &($\chi^2/dof$)\\
	    \hline
	    9670 & $\kappa$ (non-dips) & 1.77 $\pm$ 0.10 &6$\pm$2 &2.34 $\pm$ 0.18 & 1.49$\pm$1.1&7.7 $\pm$ 0.2 & 2.89 $\pm$ 0.18 &583/550& 2.87$\pm$0.64&0.35$\pm$0.07& 2.54$_{-0.23}^{+0.22}$&4.1$\pm$0.1&576/544\\
	    9670 & $\kappa$ (dips) & 1.99 $\pm$ 0.06&5.6$\pm$0.9 & 2.14 $\pm$ 0.47 &$0.17^{+0.39}_{-0.14}$& 7.5 $\pm $ 0.2 & 2.13 $\pm$ 0.11&557/528&6.69$_{-1.66}^{+0.15}$&0.10$\pm$0.01&3.40$_{-0.25}^{+0.16}$&18.9$\pm$0.4&586/529\\
	    10394 & $\omega$ (non-dips) & 1.78 $\pm$ 0.09 &10$\pm$1& 2.53 $\pm$0.22 &6$\pm$3& 7.7 $\pm$ 0.2 & 2.5 $\pm$ 0.1 &723/627&2.93$_{-0.59}^{+0.59}$&0.59$\pm$0.12&2.27$_{-0.47}^{+0.16}$&3.4$\pm$0.1&657/610\\
	    10394 & $\omega$ (dips) & 2.06 $\pm$ 0.17 & 6$\pm$2&2.54 $\pm$ 0.25 &1.3$\pm$0.1& 7.5 $\pm$ 0.3 & 2.5 $\pm$ 0.1 &436/443&5.35$\pm$0.69&0.06$\pm$0.01&$3.50_{-0.24}^{+0.08}$&42$\pm$17&429/442 \\  
	    \hline	
	    \end{tabular}
    }
    
    	\label{tab:acc_spectra}
    	
\end{table*}

\begin{figure*}
    \includegraphics[width=0.48\textwidth]{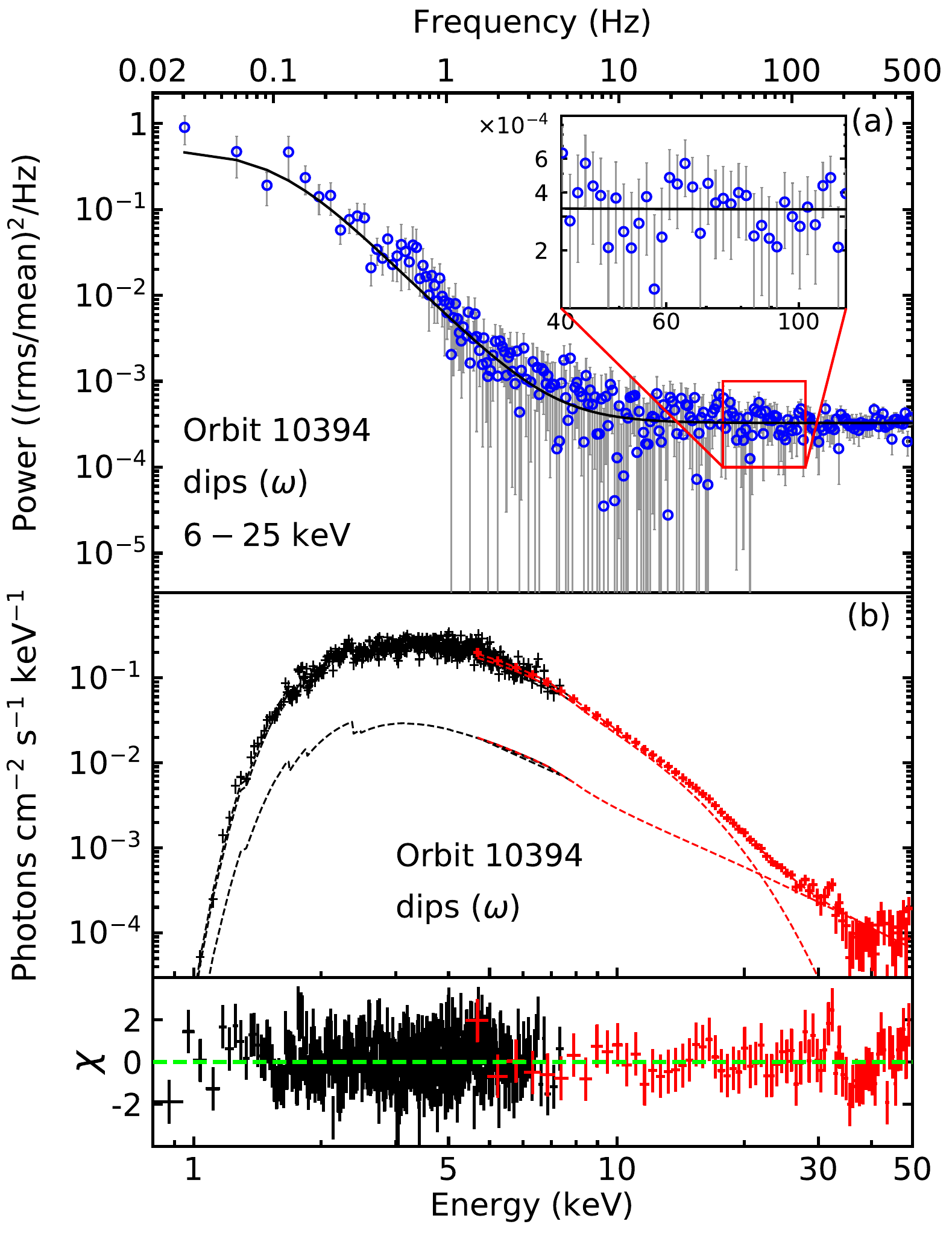}\
    \includegraphics[width=0.48\textwidth]{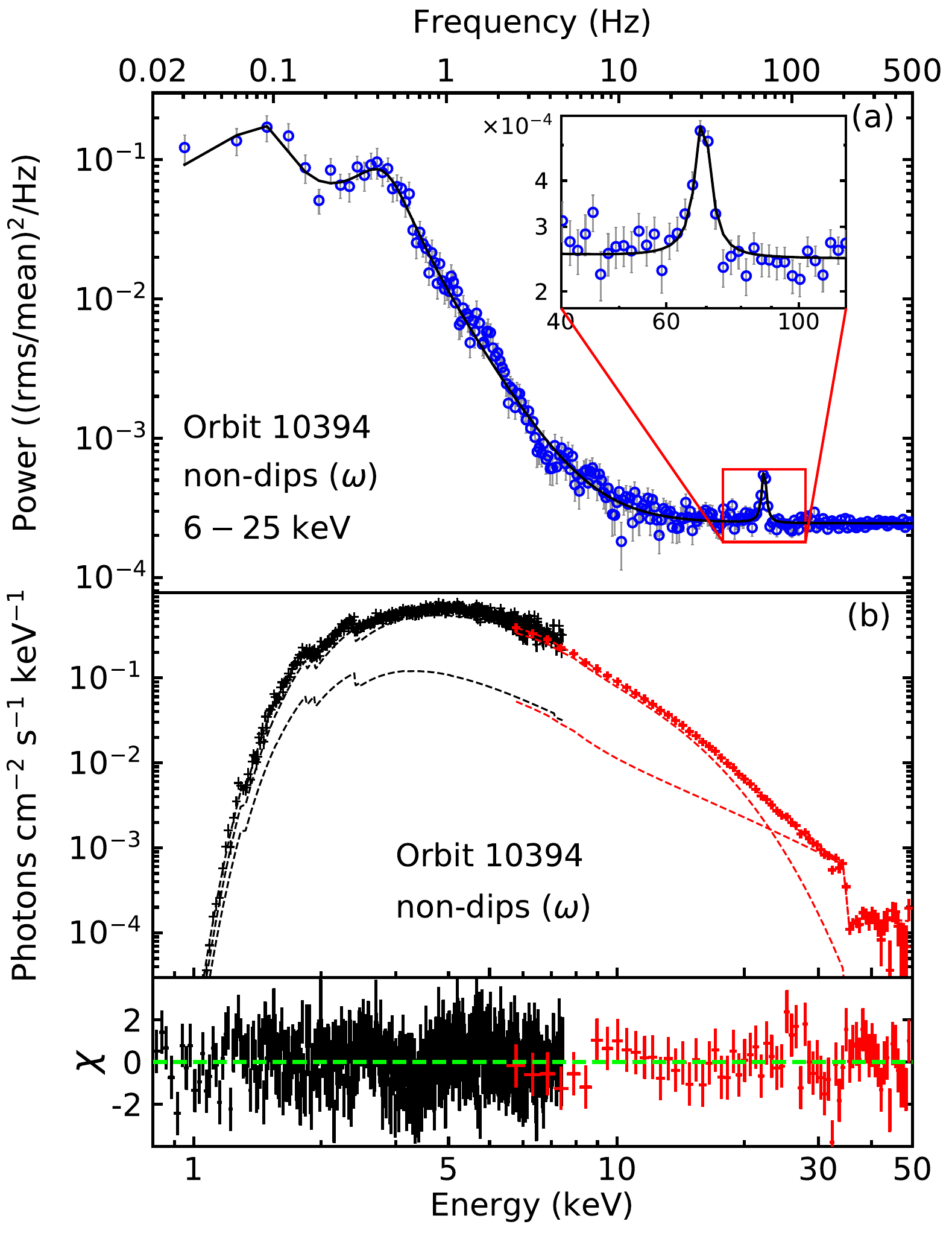}
    \caption{Left: Power density spectrum (PDS) and energy spectral distribution obtained by co-adding data from the `dips' segments of the light curve in the $\omega$ class (orbit 10394) of GRS $1915+105$ are shown in the top and bottom panels, respectively. Right: Corresponding results obtained from the `non-dips' segments of the light curve. See the text for details.}
	\label{fig:dip_non-dip_nth}
\end{figure*}

\begin{figure*}
    \includegraphics[width=\textwidth]{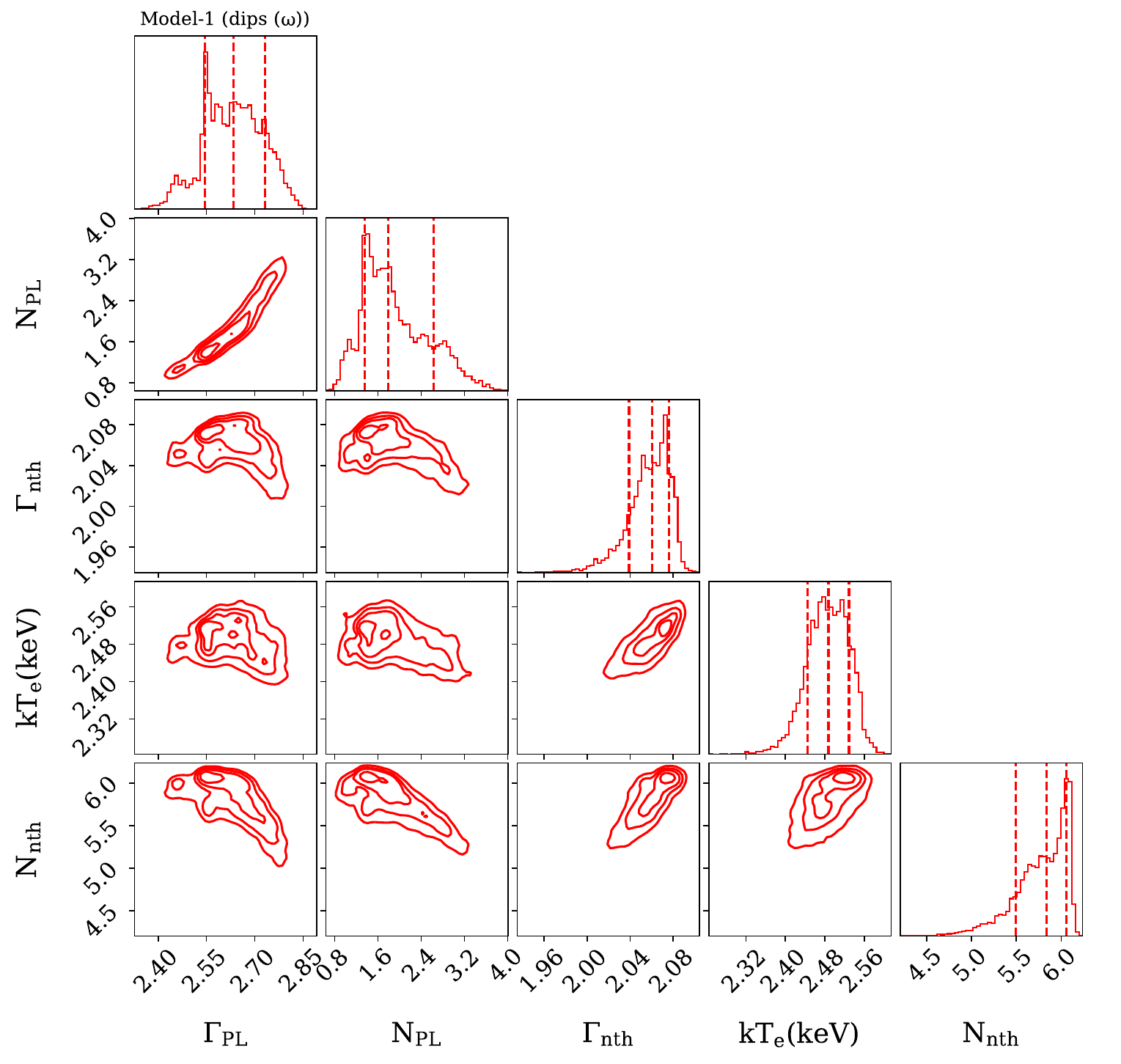}
    \caption{Correlations between parameters of \texttt{nthComp} and \texttt{powerlaw} components of Model-1 are presented for the `dips' segment of the $\omega$ class variability. In addition to the contours, we also show the distribution of each parameter on the top. The \texttt{powerlaw} parameters, such as $\Gamma_{\rm PL}$ and ${\rm N_{PL}}$, are positively correlated and both of them are anti-correlated with the \texttt{nthComp} parameters ($\Gamma_{\rm nth}$, ${\rm kT_e}$ and ${\rm N_{nth}}$). See the text for details.}
    \label{fig:contours1a}
\end{figure*}

\begin{figure*}
	\includegraphics[width=\textwidth]{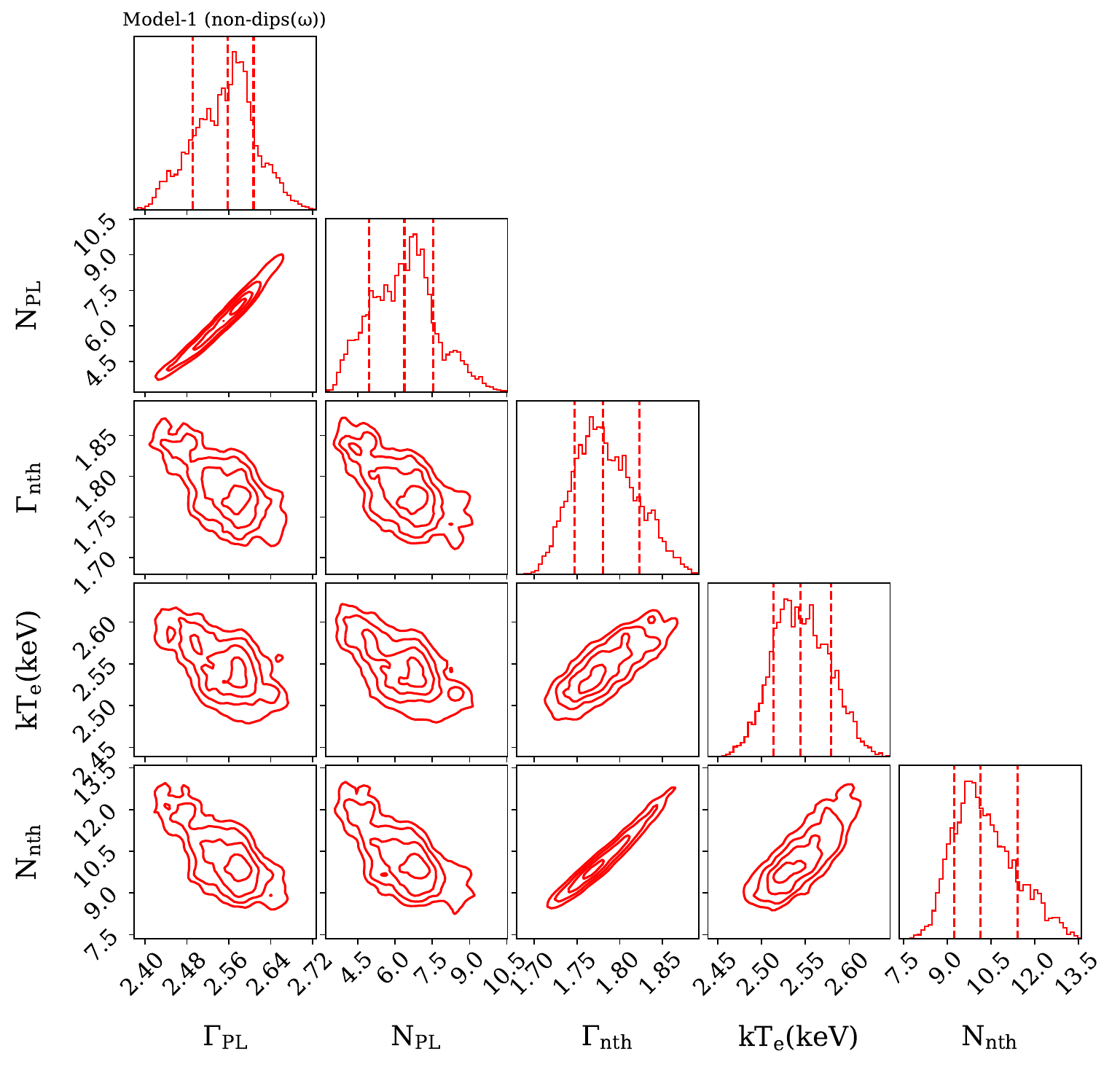}
    \caption{Same as Figure \ref{fig:contours1a}, but for `non-dips' segments. See the text for details.}
    \label{fig:contours1b}
\end{figure*}

\section{Modelling and Results}
\label{s:DaAn}

\subsection{Static Imprint of Variability Classes}

In this work, we use {\it AstroSat/LAXPC} observations to investigate the origin and characteristics of $\sim 70$ Hz HFQPOs detected in GRS 1915+105. We identify four variability classes ($i.e.$, $\delta$, $\kappa$, $\omega$, and $\gamma$ classes), where HFQPOs are observed \cite[]{Sreehari-etal2020,Majumder2022}. In Figs. \ref{fig:PDS10394}a-b, we present the MAXI light curve ($2-20$ keV) and the hardness ratio (HR: $6-20$ keV/$2-6$ keV) of GRS 1915 + 105 during {\it AstroSat} observations. The HR variation evidently indicates that the source remains in the `softer' spectral states. The vertical dashed lines denote different variability classes, and the corresponding light curves  are presented at the inset of Figure \ref{fig:PDS10394}a.

We generate deadtime corrected and Poisson noise subtracted PDS \cite[see ][for details]{Agrawal-etal2018,Sreehari-etal2020} of GRS 1915$+$105. In Figure \ref{fig:PDS10394}c, we depict PDS of the source for $\omega$ class (see Table \ref{tab:spectra}) in wide frequency band ($0.01-500$ Hz). The HFQPO feature is modelled using a \texttt{Lorentzian} with a centroid frequency of $68.09$ Hz, a Q factor ($\nu/\Delta \nu$) of $12.26$ and a significance ($\sigma={\rm norm}/{\rm err}_{\rm neg} $) of $5.62$. The insets show the PDS across different energy ranges, revealing that HFQPO is detected only in the $6 - 25$ keV range, without detection of HFQPO in the $3 - 6$ keV and $25 - 60$ keV bands \cite[see also][]{Sreehari-etal2020,Majumder2022}.

Subsequently, we perform broadband spectral modelling combining both {\it SXT} \cite[]{Singh2017} data ($0.7-7$ keV) and {\it LAXPC20} ($3-50$ keV) data for all variability classes under consideration. Initially, the broadband spectra are modelled using phenomenological model combination of \texttt{constant$\ast$Tbabs$\ast$edge$\ast$smedge$\ast$(diskbb+powerlaw)}. The \texttt{edge} component corresponds to Xenon absorption around $29-32$ keV \cite[]{Sreehari-etal2020,Majumder2022}, whereas the \texttt{smedge} component accounts for an absorption feature around 8 keV which primarily depends on the disc inclination and ionization parameters \citep{Ross1996}.
We use \texttt{gain fit} command to take care of residuals due to instrumental features at $1.8$ keV and $2.2$ keV of {\it SXT}. The best fit modelling yields the \texttt{diskbb} inner disc temperature $k T_{\rm in} \sim 2.72$ keV and \texttt{powerlaw} photon index $\Gamma \sim 2.73$.

Further, we model the broadband energy spectra using physical model components to examine the characteristics of the thermally Comptonized emissions. Accordingly, we adopt a model combination \texttt{Tbabs$\ast$edge$\ast$smedge(nthComp+powerlaw)} (hereafter Model-1), where we fix the disc black body temperature as $0.1$ keV for seed photons. The unfolded energy spectrum along with residuals of $\omega$ class is shown in Figure \ref{fig:PDS10394}d, where the energy range of dominant Comptonization is highlighted using grey shade. We find that the photon index of the thermal Comptonization component ranges as $\Gamma_{\rm nth} \sim 1.68 - 1.92$, electron temperature $kT_{\rm e} \sim 2.41 - 2.62$ keV and \texttt{powerlaw} photon index $\Gamma_{\rm PL} \sim 2.14 - 3.15$. The model fitted parameters are tabulated in Table \ref{tab:spectra}. In case of $\omega$ class (Orbit 10394), the broadband fit of \texttt{nthComp} model without the additional \texttt{powerlaw} resulted in a reduced chi-squared of $1328/627=2.11$. After adding the \texttt{powerlaw}, the reduced chi-squared improved to $724/625=1.15$ justifying the need for the \texttt{powerlaw} component. This additional \texttt{powerlaw} indicates the presence of an extended corona besides the thermally Comptonized cloud represented by \texttt{nthComp}. The relatively higher values of photon-index ($\sim 2.14 - 3.15$) of this \texttt{powerlaw} component can be attributed to non-thermal processes such as bulk motion Comptonization \citep{Nied2006,Titarchuk2009}. It may be noted that without the \texttt{smedge} (\texttt{edge}) component, Model-1 resulted in a fit with a $\chi_{\rm red}^{2}$ of $1.46 (1.44)$, justifying the inclusion of these absorption model components.
\begin{figure*}
	\begin{center}
		\includegraphics[width=0.49\textwidth]{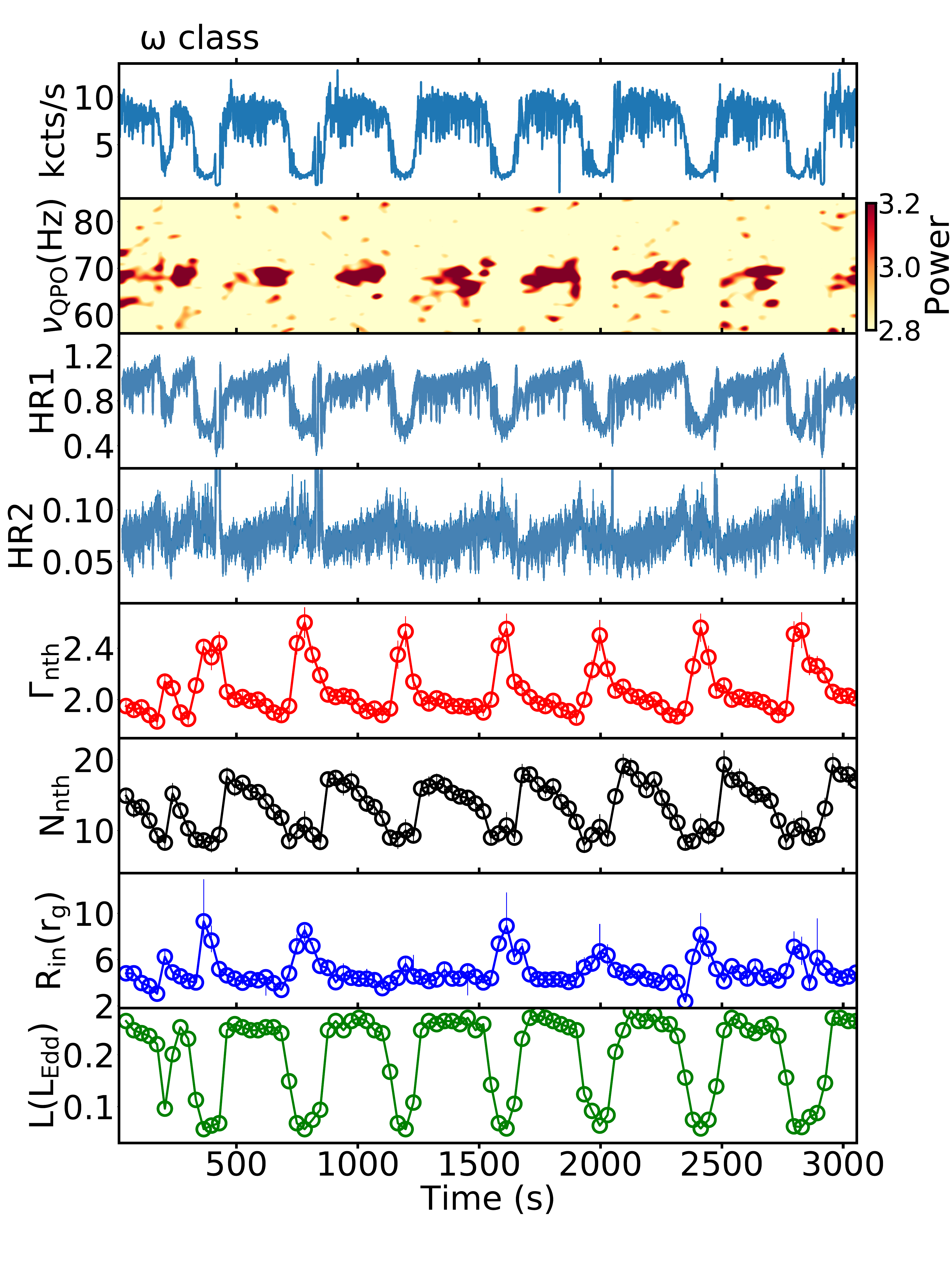}
		\includegraphics[width=0.49\textwidth]{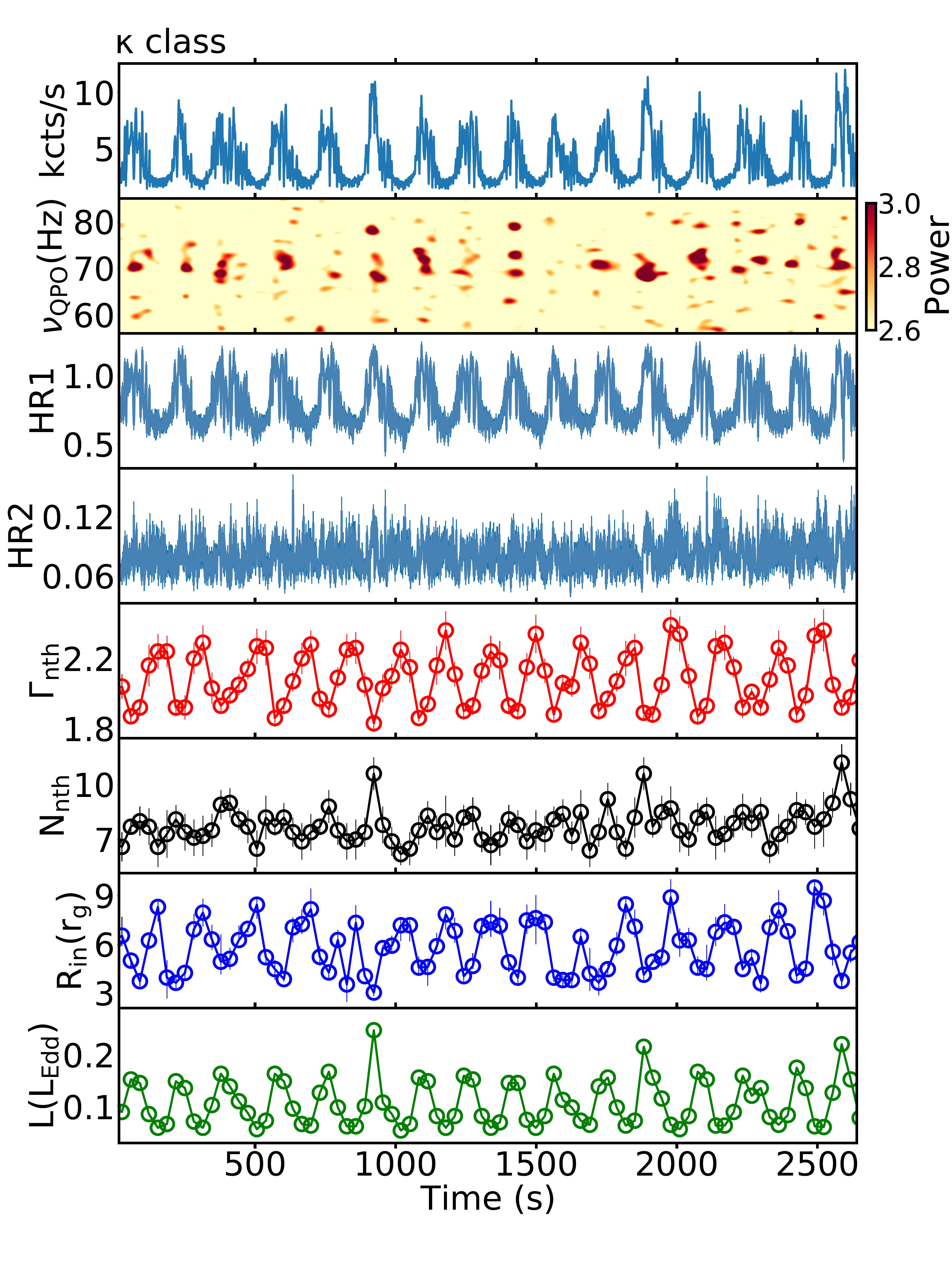}
	\end{center}
	\caption{Dynamical variation of spectro-temporal parameters for $\omega$ class ({\it left}) and $\kappa$ class ({\it right}) are shown. The count rate (kcts/s), $\nu_{\rm QPO}$ (Hz), HR1, HR2, $\Gamma_{\rm nth}$, ${\rm N}_{\rm nth}$, ${\rm R}_{\rm in}(r_{\rm g})$ and ${\rm L}~({\rm L}_{\rm Edd})$ are presented successively from top to bottom panels for $\omega$ and $\kappa$ classes. The spectral parameters are derived from Model-1 for $\Gamma_{\rm nth}$ and ${\rm N}_{\rm nth}$, and from Model-3 for ${\rm R}_{\rm in}$. Here, $\Gamma_{\rm nth}$ and ${\rm R}_{\rm in}$ are anti-correlated with count rate, meanwhile all other quantities except HR2 are correlated with count rate. The dynamic PDS in the second panel has a frequency binning of 1 Hz and it uses `bicubic' interpolation for visualization. See the text for details.}
	\label{fig:Dyn9670}
\end{figure*}

We continue modelling the source spectra replacing \texttt{diskbb} by \texttt{diskpn} \cite[]{Gierlinski1999} using model combination \texttt{Tbabs$\ast$edge$\ast$smedge(diskpn+powerlaw)} (hereafter Model-2) to estimate the size of the corona. Considering distance, mass and inclination of the source as $8.6$ kpc, $12.4$ ${\rm M}_{\odot}$ \cite[]{Reid2014,Sreehari-etal2020}, and $65^{\circ}$ \cite[]{Zd2014}, the inner disc radius is computed as ${\rm R}_{\rm in} \sim 14.62-16.56~r_{\rm g}$, where $r_{\rm g}$ is the gravitational radius. To examine the relativistic effects, we use  \texttt{kerrd} \cite[]{Ebisawa2003} model instead of \texttt{diskbb} as \texttt{constant$\ast$Tbabs$\ast$edge$\ast$smedge(kerrd+powerlaw)} (hereafter Model-3) while computing ${\rm R}_{\rm in}$. The source parameters such as distance, mass and inclination are fixed in Model-3 to the above mentioned values. Since \texttt{kerrd} model is developed considering black hole of spin $a_{\rm k} = 0.998$, we obtain the inner disc radius very close to the horizon as ${\rm R}_{\rm in} \sim 3.27 - 3.72~r_{\rm g}$. 
Based on the findings from Model-2 and Model-3, we infer that the inner disc radius (equivalently size of the corona) seems to be localized in the range $3~ r_{\rm g} \lesssim {\rm R}_{\rm in} \lesssim 16~ r_{\rm g}$. The best fit parameters for both models are presented in Table \ref{tab:spectra}. In this study, we exclude Model-2 from further analysis as it uses the pseudo-Newtonian approach to approximate gravitational effects, rather than using the general relativistic prescription. However, we utilize Model-2 solely to estimate the maximum inner disc radius that suggests the largest plausible corona geometry. 

Further, to examine the variations in spectral parameters during `dips' and `non-dips' independently, we generate separate broadband spectra combining good time intervals (gti) of only the `dips' (`non-dips') for simultaneous {\it LAXPC} and {\it SXT} observations. This analysis is carried out independently for $\omega$ and $\kappa$ class observations only as they exhibit distinct `dips' and `non-dips' in the light curve unlike other classes. The resulting broadband spectra are shown in Figure \ref{fig:dip_non-dip_nth} and the best fitted spectral parameters with Model-1 are tabulated in Table \ref{tab:acc_spectra}. N$_{\rm nth}$ and N$_{\rm PL}$ denote the normalization of the \texttt{nthComp} and \texttt{powerlaw} components respectively. We observe distinct variations in spectral parameters during `dips' and `non-dips'. For instance, the photon index of thermal comptonization remains harder ($\Gamma_{\rm nth} \sim 1.78$) during `non-dips' and it steepens ($\Gamma_{\rm nth} \sim 2.06$) during `dips' indicating softening of spectra. We present the results of Model-3 in Table \ref{tab:acc_spectra}. While $\Gamma$ is in the range $2.14-2.54$, the inner disc radius R$_{\rm in}$ is low ($2.87-2.93 r_{\rm g}$) for `non-dips' compared to `dips' ($5.35-6.69 r_{\rm g}$). We explore these effects further with our dynamic spectral analysis using {\it LAXPC} data in \S \ref{sec:Dyn}. 

In addition, we generate the MCMC (Markov Chain Monte Carlo) chains associated with the best fitted spectral parameters of `dips' and `non-dips' spectra in XSPEC. Contour plots were produced from these chains to check the correlations between spectral parameters, which are presented in Figure \ref{fig:contours1a} and Figure \ref{fig:contours1b} for $\omega$ class (Orbit 10394) observations for `dips' and `non-dips' segments. In Model-1, the spectrum consists of two additive components, namely \texttt{powerlaw} and \texttt{nthComp}.
During the `dips' (see Figure \ref{fig:contours1a}), the \texttt{powerlaw} parameters ($\Gamma_{\rm PL}$ and ${\rm N_{PL}}$) exhibit a positive correlation with each other, while both show an anti-correlation  with the \texttt{nthComp} parameters ($\Gamma_{\rm nth}$, ${\rm kT_e}$, and ${\rm N_{nth}}$). Furthermore, the \texttt{nthComp} parameters are themselves mutually correlated. A similar pattern of correlations among these parameters is observed during the `non-dips' segments as well (see Figure \ref{fig:contours1b}). However, notable difference between the spectra are observed with `non-dips' exhibiting harder spectral distribution (lower $\Gamma_{\rm nth}$) and higher normalizations for both \texttt{nthComp} and \texttt{powerlaw} components (${\rm N_{nth}}$ and ${\rm N_{PL}}$), respectively.

\subsection{Dynamical Spectro-temporal Properties}
\label{sec:Dyn}
Exploring the dynamic nature of structured variability classes enhances our understanding of the origins of HFQPOs. Following \cite{Majumder2022}, we generate dynamic power spectra using \textsc{stingray}\footnote{\url{https://pypi.org/project/stingray/}} package  \cite[]{Huppenkothen2019}, where the span of time segments is considered as $32$ s and the frequency bin size is chosen as $1$ Hz. We apply `bicubic' interpolation method to smoothen out the PDS for plotting, as it is less noisy compared to other options such as `sinc', and `spline'. `Bicubic' interpolation uses a weighted average approach by fitting cubic polynomials to 16 neighbouring pixels that form a 4$\times$4 square around the pixel under consideration to compute its value, thus reducing artifacts such as pixelation or jagged edges. The obtained results are presented in the second panels of Figure \ref{fig:Dyn9670} for $\omega$ class (left side) and $\kappa$ class (right side). The color-coded dynamic PDS indicates that dark red represents a strong signal, gradually fading to yellow as the signal strength decreases. The corresponding light curves from $\omega$ and $\kappa$ classes are shown in the top panels of Figure \ref{fig:Dyn9670}. It is evident from these light curves that there is significant oscillation in the high count-rate regions as opposed to the low count-rate regions. In the third and fourth panels of Figure \ref{fig:Dyn9670}, we present the hardness ratios HR1 and HR2, where HR1 is defined as the ratio of flux in $6 - 15$ keV to $3 - 6$ keV, whereas HR2 is the ratio of flux in $15 - 60$ keV to $3 - 6$ keV. It is evident from the figure that HFQPOs are generally detected when the photon count rates are higher (`non-dips') in both $\omega$ and $\kappa$ variability classes. Further, HR1 is also seen to correlate with the photon count rates, however no such correlation is observed for HR2.

We further examine the time-dependent behavior of the energy spectrum. In doing so, we model the energy spectrum in $3 - 25$ keV energy range using Model-1 for each successive $32$ s time segment within the {\it AstroSat} observations. Here, we restrict the energy range upto $25$ keV in order to avoid poor photon statistics at higher energies and hence \texttt{powerlaw} component is excluded while model fitting. The results are illustrated in $5^{th}$ and $6^{th}$ panels of Figure \ref{fig:Dyn9670}, depicting the time evolution of photon index ($\Gamma_{\rm nth}$) and normalization (${\rm N}_{\rm nth}$) from the \texttt{nthComp} component of Model-1. We observe that $\Gamma_{\rm nth}$ generally exhibits an anti-correlation with the photon count rate, while ${\rm N}_{\rm nth}$ aligns with the variations in the count rate. We also note that HFQPOs disappear when $\Gamma_{\rm nth}$ is higher. Notably, when HFQPOs are detected, ${\rm R}_{\rm in}$ estimated from the \texttt{kerrd} component of Model-3 is found to extend at relatively smaller radii (see $7^{th}$ panel of Figure \ref{fig:Dyn9670}) compared to the scenario when HFQPOs are absent. Similar variation of ${\rm R}_{\rm in}$ w.r.t. count rate variations were reported by \cite{Migliari2003} though it was not in the context of HFQPOs. The comparatively larger uncertainty in radius estimates during dips has to do with low photon statistics in the spectrum corresponding to that 32 s interval. This observation highlights that HFQPO features are directly associated with higher HR1 values and lower $\Gamma_{\rm nth}$, indicating relatively harder spectral states. It is worth noting that the count rates for the $\delta$ and $\gamma$ classes remain consistently high, resulting in stable spectral parameters \cite[see][]{Sreehari-etal2020, Majumder2022}.

Next, we estimate the source bolometric luminosity (${\rm L}_{\rm bol}$) in $1-100$ keV energy range as ${\rm L}_{\rm bol}=4 \pi {\rm d}^2 {\rm F}$, where ${\rm d} ~(=8.6$ kpc) refers to the source distance and ${\rm F}$ denotes the X-ray flux. For $\omega$ class observation (Orbit 10394), we obtain ${\rm L}_{\rm bol}=0.19~{\rm L}_{\rm Edd}$, where ${\rm L}_{\rm Edd} ~(=1.26 \times 10^{38} ({\rm M_{\rm BH}/M_{\odot}})$ ergs/s) is the Eddington Luminosity and ${\rm  M}_{\rm BH}~(=12.4 ~{\rm M_\odot})$ is the black hole mass. Further, we independently calculate the bolometric luminosity by adding (a) all the dip segments (low count; ${\rm L}_{\rm bol}^{\rm D}$) and (b) all the non-dip segments ({\rm high count}; ${\rm L}_{\rm bol}^{\rm N}$) of the structured variability classes ($\omega$ and $\kappa$). For $\omega$ class, we obtain ${\rm L}_{\rm bol}^{\rm D} = 0.09~{\rm L}_{\rm Edd}$ and ${\rm L}_{\rm bol}^{\rm N}=0.26~{\rm L}_{\rm Edd}$, whereas for $\kappa$ class, ${\rm L}_{\rm bol}^{\rm D} = 0.07~{\rm L}_{\rm Edd}$ and ${\rm L}_{\rm bol}^{\rm N}=0.15~{\rm L}_{\rm Edd}$. With this, we infer that HFQPOs seem to be associated with the relatively high source luminosity (see bottom panels of Figure \ref{fig:Dyn9670}). We also calculate ${\rm L}_{\rm bol}$ for $\delta$ class (Orbit 3819) and $\gamma$ class (Orbit 10583) observation. All the obtained results are presented in Table \ref{tab:spectra}.

\section{Discussion and Conclusion}
\label{s:Disc}

In this paper, for the first time, we trace the dynamical evolution of timing and spectral properties of GRS 1915+105 in small time segments ($32$ s) of the structured variability classes exhibiting $\sim 70$ Hz HFQPO feature. We use {\it AstroSat} observations in the softer variability classes, namely $\delta$, $\kappa$, $\omega$, and $\gamma$ classes. We find that for $\omega$ and $\kappa$ classes, power of $\nu_{_{\rm QPO}}$, HR1 and ${\rm L}_{\rm bol}$ correlate with the high count rates (`non-dips'), while $\Gamma_{\rm nth}$ and ${\rm R}_{\rm in}$ generally show anti-correlation. 

Wide band ($0.7-50$ keV) spectral modelling of the `softer' variability classes of GRS 1915+105 indicates the presence of both thermal Comptonization component ($\Gamma_{\rm nth} \sim 1.68-1.92$) and \texttt{powerlaw} component ($\Gamma_{\rm PL} \sim 2.14-3.15$) \cite[see][for more details]{Sreehari-etal2020,Majumder2022}. These findings seem to mimic the existence of two Comptonized components as reported by \cite{Titarchuk2009}.
To explore the variations further, we considered $\omega$ and $\kappa$ classes where the flux alters between intervals of `dips' and `non-dips'. All the `dips' segments were combined, resulting in a broadband spectrum with a photon index of $\Gamma_{\rm nth} \sim 2.06$. In comparison, the accumulated `non-dips' segments yield a comparatively harder spectrum with $\Gamma_{\rm nth} \sim 1.78$ for the $\omega$ class. The \texttt{powerlaw} index remains same $\sim 2.5$ for both `dips' and `non-dips' segments. With Model-3, we obtained an inner disc radius $< 3 ~{\rm r_g}$ for `non-dips' and $\sim 6 {\rm r_g}$ for `dips'. We associate the inner disc radius with the size of a Compton corona that is possibly present between the black hole and the disc. Contour plots of Model-1 (see Figures \ref{fig:contours1a} and \ref{fig:contours1b}) indicate that the \texttt{powerlaw} parameters are mutually correlated with each other, while they are anti-correlated with the \texttt{nthComp} parameters. It is worth mentioning that though `dips' and `non-dips' parameters show similar trends of correlation, the `non-dips' spectra are harder (lower $\Gamma_{\rm nth}$) and have higher normalization (N$_{\rm nth}$).

Notably, from a dynamic analyses of the {\it LAXPC} (3$-$25 keV) energy spectra (divided into $32$ s segments within each observation duration), we infer that the source spectral nature is softer ($\Gamma_{\rm nth} \gtrsim 2.2$, HR1 $\lesssim 0.7$) for the `dips' segments, whereas for `non-dips' segments, the spectra appear to be harder ($\Gamma_{\rm nth} \lesssim 2$, HR1 $\gtrsim 1$) (see Figure \ref{fig:Dyn9670}). Moreover, we observe that the strength of HFQPOs are directly correlated with the photon count rates. It is important to note that the dynamic energy spectra are modelled using the \texttt{nthComp} component alone, as high-quality spectral data with sufficient photon counts in short durations are available only up to $25$ keV. We binned the data at $32$ s to facilitate the analysis of dynamics at shorter time scales. This enables us to estimate the contribution from thermal Comptonization without \texttt{powerlaw} component. Evidently, the existence of HFQPOs is associated with the regions that inverse Comptonizes the soft photons to produce high energy radiations (see also Figure \ref{fig:PDS10394}d). Interestingly, $\delta$ and $\gamma$ classes do not show any structural variability in the light curves, however, their luminosities (${\rm L}_{\rm bol}$) are comparable with ${\rm L}^{\rm N}_{\rm bol}$ of $\omega$ and $\kappa$ classes (see Table \ref{tab:spectra}). 

Meanwhile, several models were proposed to explain the HFQPOs observed from GRS 1915+105. \cite{Cui-etal98,Merloni-etal99} proposed the relativistic precession model (RPM), where the orbital plane of the accreting particles do not align with the equatorial plane of the rotating black hole due to the relativistic frame-dragging effects. \cite{Morgan-etal97,Nowak-etal97} also attempted to explain the origin of HFQPOs in GRS 1915+105 based on the `diskoseismic' g-mode oscillation originally introduced by \cite{Kato-Fukue1980} for Newtonian discs. Further, \cite{Kato2004} proposed that HFQPOs in GRS 1915+105 may result from resonances between disc oscillation modes and internal disc warps. Following this, \cite{Kotrlova2020} investigated the influence of non-geodesic forces, such as magnetic fields and radiation pressure on epicyclic frequencies, concluding that several models are incompatible with rapidly spinning systems like GRO J1655-40 and GRS 1915+105. Their findings favor pressure-supported disc oscillations as a viable explanation for HFQPOs. More recently, \cite{Musoke2023} employed 3D GRMHD simulations to associate HFQPOs with radial epicyclic oscillations of a dense gas ring near the tearing radius, demonstrating that disc tearing and precession can plausibly generate a higher frequency QPO feature. Moreover, \cite{Dhaka-etal2023} reported the possible origin of QPO based on the relativistic dynamical frequency model. Meanwhile, \cite{Dihingia-etal2019} indicates that the modulation of the compact inner disc (equivalently Comptonized corona) successfully exhibits the quasi-periodic variations of the emergent flux from a viscous accretion flow.

In addition, there were attempts to explain the structured variability classes ($\omega$, $\kappa$, $\lambda$, and $\rho$), characterized by recurring `dip' and `non-dip' segments, in the light curves of GRS 1915+105 using accretion instability models. These models attribute the variability to rapid evacuation and subsequent gradual refilling of the inner accretion disc driven by viscous-thermal instabilities \cite[]{Belloni1997, Neilsen2011, Vincentelli2023}. Meanwhile, \cite{Janiuk-etal2000} demonstrated that the radiation pressure-driven instability within a standard accretion disc model can account for the intricate temporal variability observed in GRS 1915$+$105. However, using a model-independent approach, \cite{Nandi-etal2001,Chakrabarti-etal2005} suggested that these changes are likely to be associated with the sub-Keplerian accretion flow, as the observed transition timescales are much shorter than the expected viscous timescale. Furthermore, GRS $1915+105$ has been observed to display variability and class transitions over a wide luminosity range ($0.7-35\%$ of $L_{\rm Edd}$) \cite[]{Athulya-etal2022,Athulya-Nandi2023}, indicating that such variability can occur without the inner disc evacuation as proposed by disc instability models, and largely independent of the accretion rate.

Overall, the present dynamic spectro-temporal analysis of $\omega$ and $\kappa$ variability classes in GRS $1915+105$ reveals a clear association between inner disc radius variations and the presence of HFQPOs. We find that HFQPOs at $\sim 70$ Hz occur only during relatively harder spectral states ($\Gamma_{\rm nth} \lesssim 2$), where the inner disc is close to the black hole ($R_{\rm in} \lesssim 4r_g$), suggesting a compact Comptonizing corona. Conversely, in softer states ($\Gamma_{\rm nth} \gtrsim 2.2$) with larger inner disc radii ($R_{\rm in} \gtrsim 7r_g$), the HFQPOs are absent. Spectral modelling using \texttt{kerrd} and \texttt{diskpn} components constrains the maximum inner disc radius as $\sim 16 r_{\rm g}$, further supporting a link between disc truncation and corona size. Collectively, the observed dynamic variations in photon index, spectral hardness, inner disc radius, and the presence or absence of HFQPOs support the interpretation that these oscillations result from aperiodic modulations in a compact, high-energy Comptonized corona. Since the observed modulations occur on timescales shorter than the viscous timescale, a sub-Keplerian accretion flow likely governs this behavior. This conjecture possibly aligns with theoretical prescription where the coronal structures and their associated variabilities are the natural consequences of sub-Keplerian accretion under appropriate physical conditions \citep{Chakrabarti-etal1995,Das-etal2014,Sukova-Janiuk2015,Debnath-etal2024}.

\section*{Acknowledgments}
The authors are thankful to the reviewer for his/her valuable suggestions and comments that helped to improve the quality of the manuscript. S. Harikesh acknowledges support from Prof. Doron Chelouche of the Department of Physics, University of Haifa through grants from the Israeli Science Foundation (ISF 2398/19, 1650/23). AN thanks GH, SAG; DD, PDMSA; and Director, URSC for encouragement and continuous support to carry out this research. This research made use of the data obtained from the {\it AstroSat} archive of the Indian Space Science Data Center (ISSDC). The authors thank the SXT-POC of TIFR and the {\it LAXPC} team of IUCAA and TIFR for providing the data extraction software for the respective instruments.

\section*{Data Availability}
Data used for this publication are currently available at the Astrobrowse ({\it AstroSat} archive) website (\url{https://astrobrowse.issdc.gov.in/astro\_archive/archive}) of the Indian Space Science Data Center (ISSDC).



\bibliographystyle{mnras}
\bibliography{references} 



%
%
%

\bsp	
\label{lastpage}
\end{document}